# Unveiling the Mechanism of Phonon-Polariton Damping in α-MoO$_3$

*Javier Taboada-Gutiérrez[‡], Yixi Zhou[†], Ana I. F. Tresguerres-Mata[§], Christian Lanza[§], Abel Martínez-Suárez[§], Gonzalo Álvarez-Pérez[§,¥], Jiahua Duan[§,¥], José Ignacio Martín[§,¥], María Vélez[§,¥], Iván Prieto[£], Adrien Bercher[‡], Jérémie Teyssier[‡], Ion Errea[∥,‡,₃], Alexey Y. Nikitin[₃,€], Javier Martín-Sánchez[§,¥],\*, Alexey B. Kuzmenko[‡],\* and Pablo Alonso-González[§,¥],\**

[‡]Department of Quantum Matter Physics, Université de Genève, 24 Quai Ernest Ansermet, CH-1211, Geneva, Switzerland

[†]Beijing Key Laboratory of Nano-Photonics and Nano-Structure (NPNS), Department of Physics, Capital Normal University, Beijing 100048, China

[§]Department of Physics, University of Oviedo, Oviedo 33006, Spain

[¥]Center of Research on Nanomaterials and Nanotechnology, CINN (CSIC-Universidad de Oviedo), El Entrego 33940, Spain

[£]Institute of Science and Technology Austria, Klosterneuburg 3400, Austria

[∥]Fisika Aplikatua Saila, Gipuzkoako Ingeniaritza Eskola, University of the Basque Country (UPV/EHU), Europa Plaza 1, 20018 Donostia/San Sebastián, Spain

[‡]Centro de Física de Materiales (CSIC-UPV/EHU), Manuel de Lardizabal Pasealekua 5, 20018 Donostia/San Sebastián, Spain

[3]Donostia International Physics Center, Manuel de Lardizabal Pasealekua 4, 20018 Donostia/San Sebastián, Spain

[e]IKERBASQUE, Basque Foundation for Science, Bilbao, 48013 Spain

Javier Taboada-Gutiérrez - Department of Quantum Matter Physics, Université de Genève, Geneva CH-1211, Switzerland

Yixi Zhou - Beijing Key Laboratory of Nano-Photonics and Nano-Structure (NPNS), Department of Physics, Capital Normal University, Beijing 100048, China

Ana I. F. Tresguerres-Mata - Department of Physics, University of Oviedo, Oviedo 33006, Spain
Christian Lanza - Department of Physics, University of Oviedo, Oviedo 33006, Spain
Abel Martínez-Suárez - Department of Physics, University of Oviedo, Oviedo 33006, Spain
Gonzalo Álvarez-Pérez - Department of Physics, University of Oviedo, Oviedo 33006, Spain and Center of Research on Nanomaterials and Nanotechnology, CINN (CSIC-Universidad de Oviedo), El Entrego 33940, Spain

Jiahua Duan - Department of Physics, University of Oviedo, Oviedo 33006, Spain and Center of Research on Nanomaterials and Nanotechnology, CINN (CSIC-Universidad de Oviedo), El Entrego 33940, Spain

José Ignacio Martín - Department of Physics, University of Oviedo, Oviedo 33006, Spain and Center of Research on Nanomaterials and Nanotechnology, CINN (CSIC-Universidad de Oviedo), El Entrego 33940, Spain

María Vélez - Department of Physics, University of Oviedo, Oviedo 33006, Spain and Center of Research on Nanomaterials and Nanotechnology, CINN (CSIC-Universidad de Oviedo), El Entrego 33940, Spain

Iván Prieto - Institute of Science and Technology Austria, Klosterneuburg 3400, Austria

Adrien Bercher - Department of Quantum Matter Physics, Université de Genève, Geneva CH-1211, Switzerland

Jérémie Teyssier - Department of Quantum Matter Physics, Université de Genève, Geneva CH-1211, Switzerland

Ion Errea - Fisika Aplikatua Saila, Gipuzkoako Ingeniaritza Eskola, University of the Basque Country (UPV/EHU), Europa Plaza 1, 20018 Donostia/San Sebastián, Spain; Centro de Física de Materiales (CSIC-UPV/EHU), Manuel de Lardizabal Pasealekua 5, 20018 Donostia/San Sebastián, Spain and Donostia International Physics Center, Manuel de Lardizabal Pasealekua 4, 20018 Donostia/San Sebastián, Spain

Alexey Y. Nikitin - Donostia International Physics Center, Manuel de Lardizabal Pasealekua 4, 20018 Donostia/San Sebastián, Spain and IKERBASQUE, Basque Foundation for Science, Bilbao, 48013 Spain

Javier Martín-Sánchez - Department of Physics, University of Oviedo, Oviedo 33006, Spain and Center of Research on Nanomaterials and Nanotechnology, CINN (CSIC-Universidad de Oviedo), El Entrego 33940, Spain

Alexey B. Kuzmenko - Department of Quantum Matter Physics, Université de Genève, Geneva CH-1211, Switzerland

Pablo Alonso-González - Department of Physics, University of Oviedo, Oviedo 33006, Spain and Center of Research on Nanomaterials and Nanotechnology, CINN (CSIC-Universidad de Oviedo), El Entrego 33940, Spain

KEYWORDS. phonon polaritons, hyperbolic materials, van der Waals materials, low temperature s-SNOM

ABSTRACT.

Phonon polaritons (PhPs) – light coupled to lattice vibrations – in the highly anisotropic polar layered material molybdenum trioxide ($\alpha$-MoO$_3$) are currently the focus of intense research efforts due to their extreme subwavelength field confinement, directional propagation and unprecedented low losses. Nevertheless, prior research has primarily concentrated on exploiting the squeezing and steering capabilities of α-MoO$_3$ PhPs, without inquiring much into the dominant microscopic mechanism that determines their long lifetimes, key for their implementation in nanophotonic applications. This study delves into the fundamental processes that govern PhP damping in α-MoO$_3$ by combining *ab initio* calculations with scattering-type scanning near-field optical microscopy (s-SNOM) and Fourier-transform infrared (FTIR) spectroscopy measurements across a broad temperature range (8 – 300 K). The remarkable agreement between our theoretical predictions and experimental observations allows us to identify third-order anharmonic phonon-phonon scattering as the main damping mechanism of α-MoO$_3$ PhPs. These findings shed light on the fundamental limits of low-loss PhPs, a crucial factor for assessing their implementation into nanophotonic devices.

INTRODUCTION.

Polaritons – hybrid light-matter excitations - hold great promises for manipulating the flow of energy at the nanoscale[1–3]. In particular, hyperbolic phonon polaritons (HPhPs) at mid-infrared (MIR) frequencies in the orthorhombic van der Waals (vdW) crystal α-MoO$_3$[4,5], are of great interest in nanophotonics. This is mainly due to their strongly directional in-plane propagation, deep-subwavelength field confinement ($\lambda_p \approx \lambda_0/50$)[6] and ultra-low losses (lifetimes $\tau \approx 8$ ps)[4], all of them key properties for the development of planar optical nanodevices, such as those based on subdiffractional focusing[6] and lensing[7]. Interestingly, these unique properties of HPhPs stem from the inherent anisotropy of the α-MoO$_3$ crystal lattice, which is translated to its optical properties, leading to narrow spectral regions called Reststrahlen Bands (RBs) where the material behaves optically as a metal only for the electric field of light along certain crystalline directions. Consequently, the elements of the anisotropic permittivity tensor (which dictates the optical response of the material) present different signs along the crystal orthogonal axes, forcing HPhPs to propagate not along all directions but in a range of angles[4].

Although the long propagation and lifetimes of HPhPs in $\alpha$-MoO$_3$ have been analyzed in recent works employing near-field optical techniques[4,5], including low temperature studies[8], the underlying damping mechanisms governing these unique properties remain to be explored from a first-principles perspective correlated to temperature-dependent experiments. Such unbiased study incorporating the impact of phonon-phonon scattering processes is required to unveil the fundamental limits of HPhPs, including the origin of the low-temperature enhancement of their lifetime observed experimentally[8].

Here, we present first principles calculations based on density functional perturbation theory, which allows us to estimate phonon lifetimes without any empirical parameter, in conjunction with an experimental cryogenic-SNOM study of the PhP propagation and lifetimes in $\alpha$-MoO$_3$ to unveil the fundamental damping mechanism of HPhPs in $\alpha$-MoO$_3$.

EXPERIMENTAL SECTION.

*Far-Field Temperature-Dependent Characterization -*

Since PhPs arise from the coupling between photons and phonons, we start by exploring how the phonon spectra in $\alpha$-MoO$_3$ vary with temperature. To do this, we carry out far-field reflectance spectroscopy on a thick $\alpha$-MoO$_3$ flake (thickness > 1 μm) using an FTIR microscope equipped with a He-flow cryostat operating down to 5 K. Top panels of Figure 1A and B show the gold-normalized reflectance spectra at 5 K (blue) and 300 K (red) for the two orthogonal in-plane polarizations of light ([100] and [001] crystallographic directions, respectively). For both polarizations and at both temperatures, we observe a high-reflectivity band from 820 to 970 cm$^{-1}$ for the [100] direction and from less than 600 cm$^{-1}$ to 851 cm$^{-1}$ for the [001] direction. Notice the limited range of the FTIR's MCT detector precludes us precisely identifying the low energy limit of this high-reflectivity band. We assign these high-reflectivity bands to the presence of RBs where HPhPs can be excited in this material based on our previous reports[4,9]. In addition, a peak is found at about 1010 cm$^{-1}$ for both in-plane polarizations, which we can relate to the presence of a third RB along the out-of-plane [010] direction (from approximately 960 to 1010 cm$^{-1}$ as detailed in Supplementary Information S3). This RB can be observed due to the finite angle of incidence of the IR light beam on the sample. It is noteworthy to stress that we observe that the RBs are slightly broader at 5 K than at 300 K, meaning that the TO and LO phonons shift spectrally with temperature.

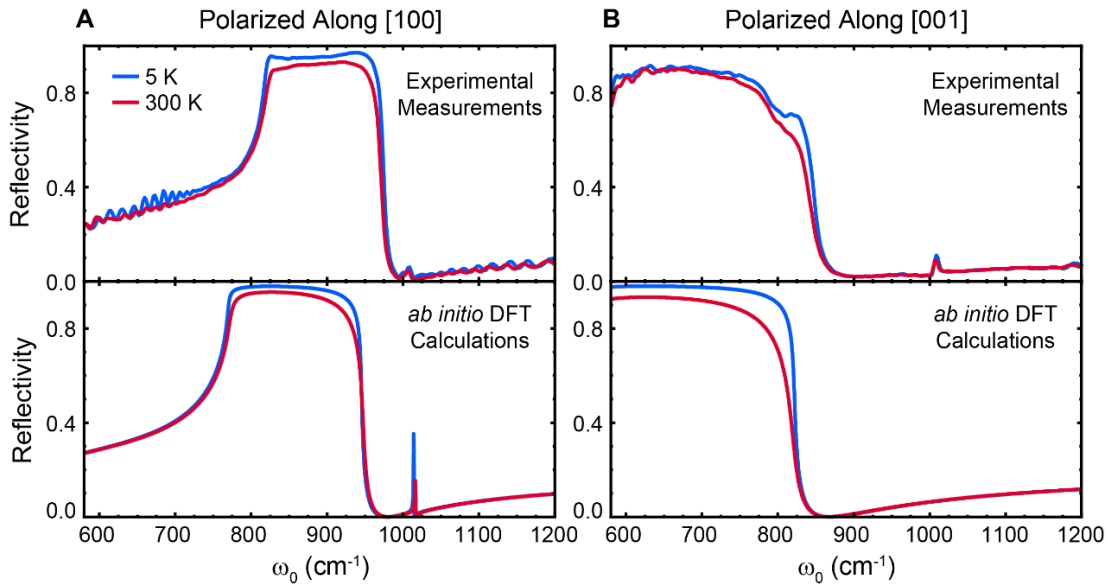

**Figure 1.** Experimental (FTIR) and theoretical (DFPT) reflectivity spectra of α-MoO$_3$. A, B Experimental (top panels) and theoretical (bottom panels) reflectivity spectra of α-MoO$_3$ for the [100] and [001] polarizations, respectively. Experimental measurements were taken by FTIR

whereas theoretical curves were calculated employing the *ab initio* extracted permittivity and supposing normal incidence of light.

To better understand these experimental FTIR spectra, we performed ab initio DFPT calculations considering semi-local exchange-correlation functionals[10]. The experimental reported lattice parameters[9] are employed with internal relaxation of the atomic positions, based on which the phonon frequencies, polarization vectors, effective charges and high-frequency dielectric constants are computed. The phonon lifetime is calculated as a function of temperature employing perturbation theory that considers three-phonon anharmonic interactions, i.e. where a phonon can exclusively decay into two other phonons. The resulting dielectric function is used to calculate the reflectance spectra. The results are shown in Figure 1A and B bottom panels, where light is supposed to illuminate the sample at a normal incidence. All computed curves qualitatively agree with the experimental ones, correctly predicting the presence of the RBs and the trends observed with temperature. We would like to highlight that the sharp structure at 1020 cm$^{-1}$ in the theoretical curve for the [100] direction (Figure 1A lower panel) is due to a weak phonon mode that competes in the experimental curve (Figure 1A top panel) with the mentioned peak that arise from the oblique incidence of light due to the presence of the [010] RB as detailed in Supplementary Information S3.

Some observed deviations between the theoretical and observed phonon frequencies are expected for a semi-local approximation of the exchange interactions and electron correlations in oxides[11]. It should be noted that only those terms which have a contribution to the phonon lifetime at the lowest order in perturbation theory have been considered, i.e. we exclusively consider the so-called anharmonic bubble self-energy diagram[12]. Therefore, the thermal expansion of the unit cell and effects related to the interaction of four or more phonons are discarded in the calculations both of the phonon lineshifts and lifetimes. The in-plane thermal expansion in this material is indeed very small[13], justifying our approximation. Impurities are also discarded as an extra scattering mechanism channel. In other words, only three-phonon interactions (two-phonon collision to form another phonon or the annihilation of a phonon to form two phonons) are considered[13]. The electron-phonon interaction is also ignored as its contribution in large-bandgap semiconductors such as α-MoO₃ ($\approx$ 3 eV) is weak. The discarded contributions may cause the phonon frequencies to shift, which explains the mentioned mismatch between the experimental and calculated spectral locations of the RBs, although the mismatch is mainly coming from the choice of the exchange-correlation functional. However, the advantage of these approximations is that temperature, which affects the thermal occupation of the phonon modes, is the only (not adjustable) parameter, allowing a stringent comparison between theoretical and experimental trends as the sample is cooled down.

To quantify experimentally the spectral shift of the α-MoO₃ TO phonon with temperature observed in the reflectivity spectra in Figure 1A top panel, we model the α-MoO₃ permittivity tensor by a sum of Drude-Lorentz oscillators[14] ($\alpha = [100], [001]$):

$$\varepsilon_\alpha(\omega) = \varepsilon_{\infty,\alpha} + \sum_{k=1}^{N} \frac{\omega_{p,k,\alpha}^2}{\omega_{TO,k,\alpha}^2 - \omega^2 - i\omega\gamma_{k,\alpha}} \quad (1)$$

Where $\varepsilon_{\infty,\alpha}$ is the high-frequency permittivity and $\omega_{p,k,\alpha}$, $\omega_{TO,k,\alpha}$ and $\gamma_{k,\alpha}$ are the 'plasma' frequency, TO frequency and the scattering rate of the k-th phonon mode, respectively. By least-square fitting the experimental data, we extract the phonon parameters at each temperature. In Figure 2A, the red symbols show the shift of the TO phonon frequency along the [100] direction, $\Delta\omega_{TO}(T) = \omega_{TO}(T) - \omega_{TO}(T_{min})$, with respect to the lowest temperature $T_{min}$ (in our case 5 K). The corresponding theoretical values for the same polarization (gray symbols) agree with the experiment remarkably well: in both cases the mode shows a softening by about 3 cm$^{-1}$ as the sample is cooled from room temperature to 5 K. This result supports the validity of our theoretical calculations and the approximations employed. Notice that the shift of the TO phonon frequency with respect to the lowest temperature is plotted instead of the TO frequency due to the frequency mismatch of the *ab initio* calculated phonon frequencies commented above.

The red symbols in Figure 2B present the experimental temperature dependence of the phonon scattering rate $\gamma$ for the [100] direction and the corresponding DFPT calculation is given by gray symbols. Despite some noise in the experimental values at low temperatures, the agreement is very satisfactory, both in terms of the absolute value and the temperature dependence (a decrease by about 2 cm$^{-1}$ across the entire cooling process). This indicates that the thermal broadening of the phonon linewidth is dominated by the third-order anharmonic phonon-phonon processes, which are affected by thermal occupation factors, in line with phenomenological results reported previously[8] and justifying our computational approximations.

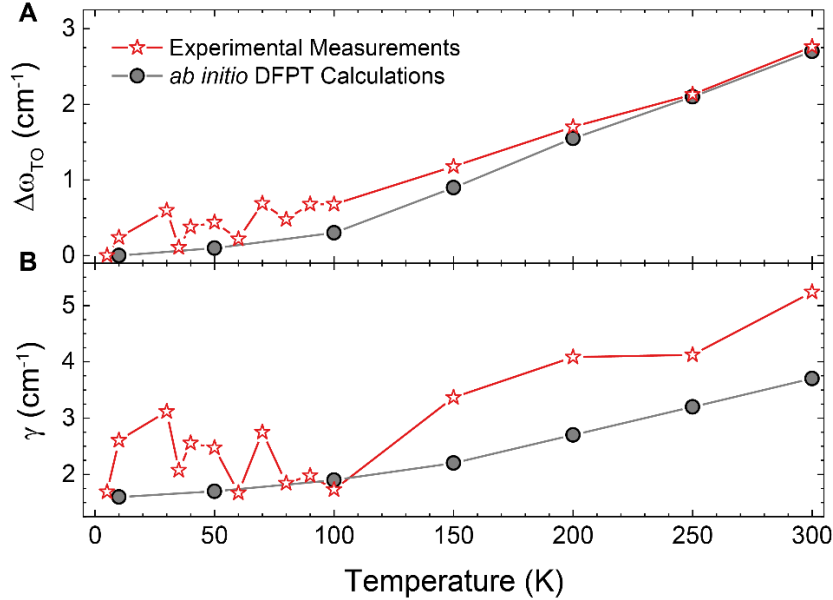

**Figure 2.** Comparison between experimental and theoretical temperature dependence of the phonon parameters in α-MoO$_3$. a) Experimental and theoretical shift of the TO phonon position $\Delta\omega_{TO}$ values showing the same trend with temperature. A hardening of 3 cm$^{-1}$ is found in both cases. b) Comparison between the experimentally extracted and theoretically calculated α-MoO$_3$ phonon linewidth, $\gamma$, for the phonon resonance along the [100] direction which gives rise to the RB between ≈ 800 cm$^{-1}$ and ≈ 1000 cm$^{-1}$.

*Near-Field Temperature-Dependent Characterization -*

For studying the damping mechanisms of PhPs in α-MoO$_3$, we perform near-field nanoimaging using a cryo-s-SNOM system (See methods and Supplementary Information S1), which allows us to directly visualize their propagation as a function of temperature. As previously reported[15], the technique is based on polariton interferometry: an oscillating atomic force microscopy (AFM) tip is illuminated with MIR light creating a dipole at the tip apex, allowing both excitation and detection of PhPs once reflected from the edges of the sample. While scanning the tip across the sample, the distance travelled by the reflected polaritons is modulated, giving rise to interferometric fringes. Both the tip and the sample are mounted in a vacuum chamber, where the temperature can vary between approximately 8 and 300 K. The s-SNOM amplitude images taken at 10, 90, 150 and 225 K on a 104 nm-thick α-MoO$_3$ flake exfoliated on SiO$_2$ are shown in Figure 3A. The illumination frequency is 880 cm$^{-1}$, for which the dielectric permittivity along the [100] direction is negative and the ones for the two other directions are positive, giving rise to hyperbolic phonon polaritons (HPhPs)[9]. In agreement with previous measurements carried out at room temperature[4], we observe oscillating fringes that are parallel to only one of the flake edges, revealing the propagation of PhPs along the [100] crystal direction. The line profiles along this crystal direction (Figure 3B) demonstrate the existence of two contributions with different periodicities[16]. Tip-excited and edge-launched PhPs can coexist in α-MoO$_3$ producing fringe oscillations with periods $\lambda_p/2$ and $\lambda_p$, respectively. Taking into account that along the [100] crystal direction the pointing vector, $S$, and the wavevector, $k_p$, are parallel[17], these profiles can be well modeled using a simple equation:

$$y(x) = y_0 + A_1 e^{-x/L_p} \sin\left(2\pi \frac{x-x_1}{\lambda_p}\right)\frac{1}{x} + A_2 e^{-2x/L_p} \sin\left(4\pi \frac{x-x_2}{\lambda_p}\right)\frac{1}{\sqrt{x}} \quad (2)$$

where $y$ is the near-field overall amplitude, $y_0$ a vertical offset, $A_1$ is the amplitude of the edge-launched wave, $A_2$ is the amplitude of the tip-launched wave, $L_p$ is the polariton propagation length, $\lambda_p$ is the polariton wavelength, and $x_1$ and $x_2$ are phase offsets, respectively. The geometrical term $1/\sqrt{x}$ considers the geometrical spreading resulting from the cylindrical nature of the tip-launched polaritons, while the factor $1/x$ is introduced to model the signal decay caused by the variation in the excitation efficiency of the edge-launched polaritons[18].

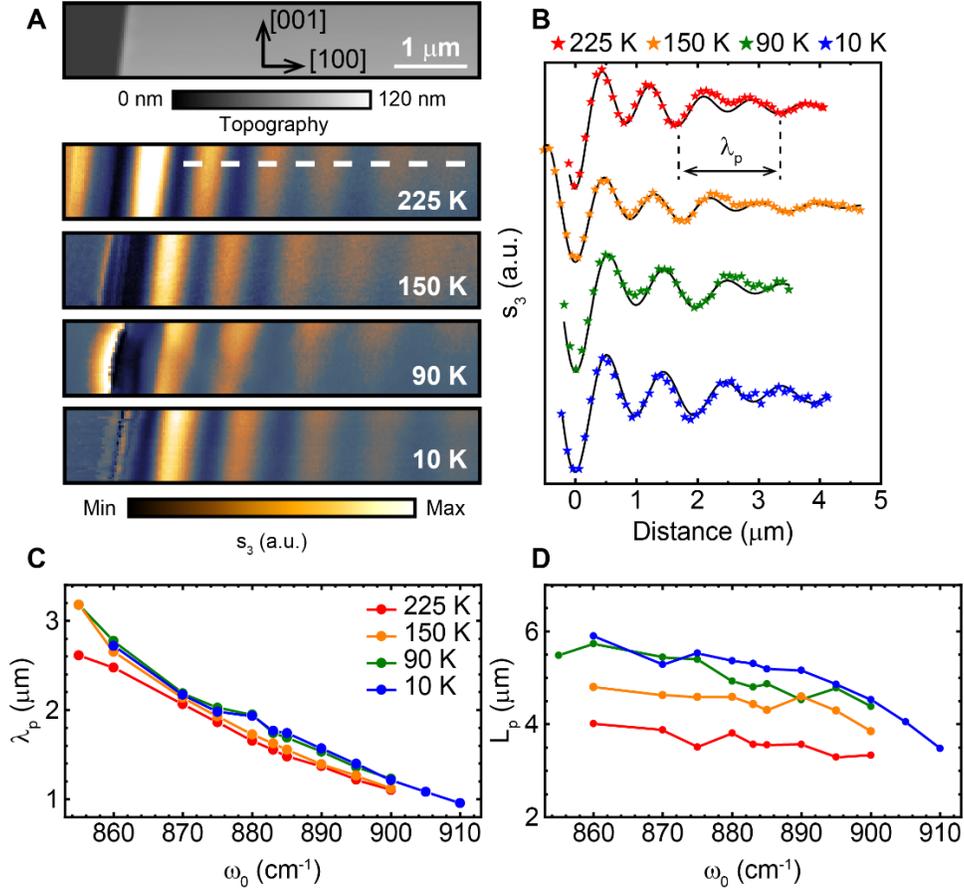

**Figure 3.** Experimental cryo-SNOM measurements of PhPs in α-MoO$_3$. a) Topography and s-SNOM images taken in a 104 nm-thick α-MoO$_3$ flake at an illuminating frequency of 880 cm$^{-1}$ and temperatures 225, 150, 90 and 10 K. The α-MoO$_3$ crystal directions are shown within the topographic image. b) Near-field amplitude profiles extracted along the white dashed lines in (a). Fittings using Equation 2 are shown as black curves. c) Experimental polaritonic wavelength $\lambda_p$ extracted from the fitting of these profiles for the hyperbolic regime at all measured temperatures and frequencies. d) Experimental polaritonic propagation length $L_p$ extracted from the fittings of the profiles shown in (b) employing Equation 2 for all temperatures and frequencies measured. A clear increase in the propagation length is found as the temperature decreases.

Figures 3C and 3D show the experimentally extracted polaritonic wavelength, $\lambda_p$, and propagation length, $L_p$, respectively, for the same temperatures and several incident frequencies from 850 to 910 cm$^{-1}$, all of them falling within the hyperbolic regime. It is noteworthy that at all temperatures we observe a consistent decrease in the HPhPs wavelengths $\lambda_p$ (Figure 3C) with increasing frequency, in agreement with previous room-temperature results[4]. Remarkably, for all frequencies, there are also slight variations of $\lambda_p$ as a function of temperature (average increase of 5% from 225 K to 10 K). This observation could constitute yet another strategy for controlling the PhPs wavelengths, albeit with modest variations. More importantly, the PhPs propagation length $L_p$ (Figure 3D) shows a distinct increase with decreasing temperature, being up to 58% at a frequency of 875 cm$^{-1}$ in the 225 K - 10 K range. This result unequivocally indicates a reduction in the phonon-phonon scattering rate at lower temperatures. These experimental findings are fully aligned with theoretical results utilizing the *ab initio* calculated permittivity (see calculations based on the dispersion relation for electromagnetic modes in biaxial slabs embedded between two isotropic media in Supplementary Information S4)[19]. However, it is important to note that a direct comparison between the absolute values of experimental and theoretical $\lambda_p$ and $L_p$ is not possible due to the slight mismatch between ab initio and experimental phonons[20]. Similar results are also observed for α-MoO$_3$ HPhPs in the elliptical regime (see Supplementary Information S4).

The polaritonic lifetime $\tau$ can be calculated according to the formula $\tau = L_p/v_g$, where $v_g$ is the PhP group velocity[4,21], which is worked out as the numerical derivative of the extracted PhP dispersions (see Supplementary Section S4). Remarkably, and in contrast to $\lambda_p$ and $L_p$, a key property of $\tau$ is that it shows a near-constant growth in frequency, and therefore as a first approximation comparing the experimental and the theoretical lifetimes is feasible regardless the frequency shifts of the *ab initio* calculated phonon positions (see Figure S9). The resulting experimental and theoretical PhPs lifetimes are shown in Figure 4 for the hyperbolic regime (Supplementary Section S5 extends this study to the elliptical regime in the frequency range 980 – 1000 cm$^{-1}$). A good agreement between experiment (star symbols) and theory (circles) is obtained. As an example, the theoretical lifetimes at 10 K are 1.86 and 2.59 ps for $\omega_0 = 860$ and 895 cm$^{-1}$, respectively, while the experimental values are $1.3 \pm 0.3$ and $2.5 \pm 0.5$ ps, respectively.

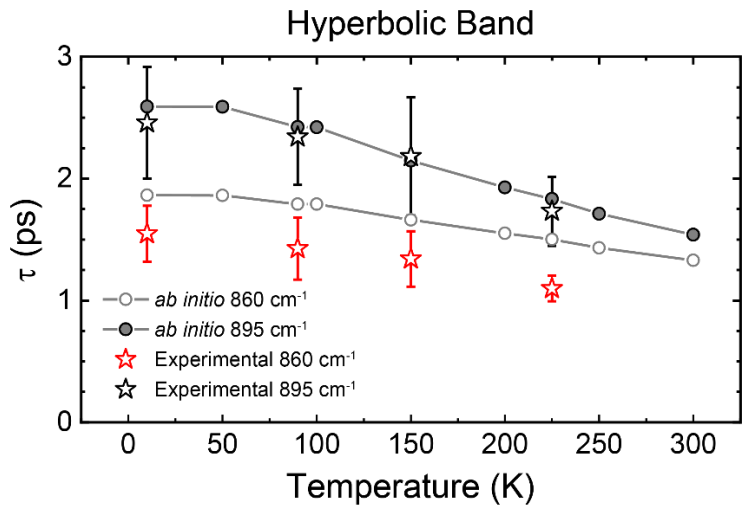

**Figure 4.** Phonon Polariton Lifetimes in α-MoO$_3$ as a Function of the Temperature. Theoretically calculated (circles) and experimentally extracted (star symbols) temperature-dependent lifetimes of PhPs for α-MoO$_3$ (104nm-thick flake) for the hyperbolic RB ($\omega_0 = 860$ cm$^{-1}$ and $\omega_0 = 895$ cm$^{-1}$). Gray straight lines serve as visual guides.

Given the good agreement between theoretical calculations and experimental observations, and considering that the theory takes into account exclusively third-order anharmonic phonon-phonon scattering (thermal expansion of the unit cell, scattering processes involving four or more phonons, crystal defects, or electron-electron and electron-phonon interactions are not considered), our findings provide a clear and unambiguous demonstration that such a mechanism constitutes the primary source of damping of PhPs in α-MoO$_3$. In fact, anharmonicity is small in this system (the phonon linewidth is two orders of magnitude smaller than the frequency) and, as it is consequently expected, the lowest-order calculation of the lifetime yields a good agreement with experiments. It should be noted that similar conclusions can also be drawn for the case of elliptical PhPs in α-MoO$_3$ (see Supplementary Section S5).

CONCLUSIONS.

To summarize, we have studied the effect of thermally induced phonon scattering processes on the propagation properties of PhPs in α-MoO$_3$ crystals. The existence of PhPs in van der Waals polar materials are intimately related to phonon resonances whose properties in terms of linewidth and spectral position strongly depend on the temperature. Our findings reveal that third order anharmonic phonon scattering processes alone explain the damping mechanisms of PhPs in α-MoO$_3$. These results are important for understanding the fundamental limits in phonon-polaritonic propagation, which is essential for pushing the boundaries of what is technologically possible in nanooptics using these excitations.

METHODS

**Near-Field Imaging at Cryogenic Temperatures:** Mid-Infrared nanoimaging was carried out by means of a commercially available (Neaspec GmbH) cryogenic scattering-type scanning near-field optical microscopy system (Cryo s-SNOM). A Pt-Ir metallized atomic force microscope (AFM) tip (NanoAndMore GmbH) is illuminated with mid-infrared light from a quantum cascade laser (QCL, Daylight Solutions Inc.) oscillating at a tip amplitude of approximately 100 nm and a tip frequency around 300 kHz allowing us to excite polaritons with a spatial resolution that depends only in the metal-coated AFM tip radius. The excited polaritons travel along the material getting reflected on the edges of the flake and travelling back along the same direction. Right after, polaritons are scattered to the far-field with the help of the AFM tip. serving thus as excitation and collection source at the same time. The incident beam is driven to a Michelson interferometer through a beam splitter which eventually mixes with the scattered field, which is finally sent to a mercury cadmium telluride (MCT) detector (Kolmar Technologies Inc.). Demodulating the registered light at high harmonics of the tip oscillating frequency ($\Omega$) permits us to obtain background-free signals with complete information about the polaritonic amplitude and phase electric field.

**Fourier-Transform Infrared Microscopy at Cryogenic Temperatures:** Mid-infrared reflectance measurements were carried out with a Bruker Vertex 70v FT-IR spectrometer attached to a Bruker HYPERION 2000 FT-IR microscope, which allows us to perform reflection, transmission and absorption measurements from the far-infrared to the near-infrared regime. Measurements were taken between 600 and 4000 cm$^{-1}$, employing an MCT detector. Measurements polarized along both in-plane directions were taken with a resolution of about 2 cm$^{-1}$. The sample was introduced into a cryostat (Cryovac Konti Micro) equipped with a zinc selenide (ZnSe) window (transparent at mid-infrared frequencies) to be cooled down until 10 K.

**Density Functional Theory Calculations:** We calculated the permittivity of $\alpha$-MoO$_3$ entirely from first principles using density functional theory (DFT). Specifically, we used the Perdew-Burke-Ernzerhof (PBE) approximation for the exchange-correlation functional[10]. Ultrasoft pseudopotentials are used in the calculations including 6 valence electrons for O and 14 valence electrons for Mo. We started with the experimentally measured crystal lattice parameters and then relax internally the atomic positions[9]. We employ perturbation theory to calculate the anisotropic permittivity of the material (which is diagonal as α-MoO₃ belongs to an orthorhombic crystal system)[22]. In this framework, the dielectric tensor for Cartesian directions $\alpha$ and $\beta$ is given as

$$\varepsilon_{\alpha\beta}(\omega) = \varepsilon_{\alpha\beta}^{\infty} + 4\pi\chi_{\alpha\beta}(\omega) \quad (3)$$

With $\varepsilon_{\alpha\beta}^{\infty}$ the high-frequency limit of the permittivity tensor and $\chi_{\alpha\beta}(\omega)$ the susceptibility, expressed in atomic units as:

$$\chi_{\alpha\beta}(\omega) = -\frac{1}{\Omega}\sum_i \frac{2\omega_i M_i^{\alpha} M_i^{\beta}}{\omega^2 - \omega_{TO,i}^2 - 2\omega_{TO,i}\Pi_i(\omega_{TO,i})} \quad (4)$$

With $\Omega$ the volume of the unit cell, $\omega_{TO,i}$ the $i$-th mode phonon frequency (the summation only considers phonon modes at the Γ point), $\Pi_i(\omega_{TO,i})$ the $i$-th phonon mode's self-energy due to phonon-phonon interaction taken at the phonon frequency, and $M_i^{\alpha}$ is the contribution of the $i$-th mode to the dipole moment:

$$M_i^{\alpha} = \sum_{\beta s} \frac{Z_s^{\alpha\beta} e_{is}^{\beta}}{\sqrt{2M_s\omega_i}} \quad (5)$$

This is, $M_i^{\alpha}$ is expressed through the effective-charge tensor for atom $s$, $Z_s^{\alpha,\beta}$, the mass of atom $s$, $M_s$, and the polarization vector of mode $i$ for atom $s$ along the Cartesian direction $\beta$, $e_{is}^{\beta}$. Phonon frequencies, polarization vectors, effective charges, and the high-frequency limit of the dielectric function are obtained by means of density functional perturbation theory (DFPT) in the way it is implemented in Quantum Espresso[20,23,24]. An 80 Ry plane-wave basis cutoff and a 800 Ry density cutoff were employed. Electronic integrations were performed on a 8x2x8 grid. The phonon self-energy was calculated only considering the bubble contribution[25,26]:

$$\Pi_i(\omega) = -\frac{1}{2N_q}\sum_{qjk}|\Phi_{ijk}(0,q,-q)|^2 \left[\frac{2(\omega_{j,q}+\omega_{k,-q})[1+n_B(\omega_{j,q})+n_B(\omega_{k,-q})]}{(\omega_{j,q}+\omega_{k,-q})^2-(\omega+i\delta)^2} + \right.$$

$$\left.\frac{2(\omega_{j,q}-\omega_{k,-q})[n_B(\omega_{k,-q})-n_B(\omega_{j,q})]}{(\omega_{k,-q}-\omega_{j,q})^2-(\omega+i\delta)^2}\right] \quad (6)$$

$\Pi_i(\omega)$ represents the lowest order term that contributes to the self-energy and has an imaginary part[12] and thus gives the correct lifetime in perturbation theory in the case of weakly anharmonic systems like $MoO_3$. The frequency of the $j$-th mode at the $q$ point of the Brillouin zone is represented by $\omega_{j,q}$, the total number of $q$ points in the summation by $N_q$, the Bose-Einstein occupation factor is represented by $n_B(\omega)$, $\delta$ accounts for a small number (10 cm$^{-1}$ in our case), and the anharmonic third-order force constants, transformed to the phonon mode basis, are given by $\Phi_{ijk}(0,q,-q)$[26]. Calculations at different temperatures are obtained changing the temperature in the Bose-Einstein occupation factor, neglecting thermal expansion. Third-order anharmonic force-constants were computed using finite-differences calculating atomic forces on displaced supercells generated by the ShengBTE code[27,28]. The calculations were performed in a 2x1x2 supercell, including interaction terms up to the 5th nearest neighbor. A 16x4x16 phonon grid included in the summation over the $q$ points is utilized for computing the phonon self-energy. By Fourier interpolation, we obtain the phonon frequencies and the third-order force constants. Initially, a grid of 6x2x6 $q$ points is employed for the calculation of the dynamical matrices. The *ab initio* scattering rates of the phonon mode $\omega_i$ reported in the main text correspond to the imaginary part of the phonon self-energy taken at the phonon frequency:

$$\gamma_i = -\text{Im}\,\Pi_i(\omega_{TO,i}) \quad (7)$$

The anharmonic shift of the frequency is calculated as:

$$\Delta\omega_{\text{TO},i} = \text{Re}\,\Pi_i(\omega_{\text{TO},i}) \quad (8)$$

We neglect the effect on the shift of the thermal expansion and the loop self-energy diagram, even if both effects are considered of the same order in perturbation theory of the loop diagram considered here.

## ASSOCIATED CONTENT

**Supporting Information**. The Supporting Information document provides the following information: S1 - Scattering-Type Scanning Near-Field Optical Microscopy (s-SNOM) and Fourier Transform Infrared Spectroscopy (FTIR). S2 - Calculated Spectra of the Dielectric Function and Reflectivity for the Normal Angle of Incidence. S3 - Effect of the Finite Angle of Incidence on the Experimental Reflectivity Spectra. S4 - Polaritonic Wavelength, Group Velocity and Propagation Length: Experiment vs. Theory. S5 - Temperature-Dependent Study of the Polaritonic Lifetime in the Elliptical Band.

## AUTHOR INFORMATION

**Corresponding Author**

*Javier Martín-Sánchez : javiermartin@uniovi.es

*Alexey B. Kuzmenko : Alexey.KuzMenko@unige.ch

*Pablo Alonso-González : pabloalonso@uniovi.es

**Author Contributions**

J.T-G., P.A-G., J.M-S. and A.B.K. conceived the study. J.T-G. performed the s-SNOM measurements with the help of Y.Z and A.B. J.T-G. and Y.Z. conducted the FTIR measurements with the help of J.T. A.I.F.T-M., A.M-S., J.I.M., M.V. and I.P. fabricated the sample. I.E. carried out the *ab initio* calculations. J.T-G. and C.L-G. performed the PhP theoretical calculations with the help of G.A-P and A.Y.N. J.T-G, J.D., C.L-G., A.B.K. and P.A-G. contributed to the data analysis. J.T-G., P.A-G. and A.B.K cowrote the manuscript with the help of all co-authors.

ACKNOWLEDGMENT

Funding Sources -

A.I.F.T.-M. and G.Á.-P. acknowledge support through the Severo Ochoa program from the Government of the Principality of Asturias (references PA-21-PF-BP20-117 and PA20-PF-BP19-053, respectively). A.B.K and J.T-G. acknowledge support from the Swiss National Science Foundation (grant # 200020_201096). J.M.-S. acknowledges financial support through the Ramón y Cajal program from the Government of Spain (RYC2018-026196-I). P.A.-G. acknowledges support from the European Research Council under starting grant no. 715496, 2DNANOPTICA and the Spanish Ministry of Science and Innovation (State Plan for Scientific and Technical Research and Innovation grant number PID2019-111156GB-I00). A.Y.N. acknowledges the Spanish Ministry of Science and Innovation (grant PID2020-115221GB-C42) and the Basque Department of Education (grant PIBA-2023-1-0007). MV and JIM acknowledge support by Spanish MCIN/AEI/10.13039/501100011033/FEDER,UE under grant PID2022-136784NB and by Asturias FICYT under grant AYUD/2021/51185 with the support of FEDER funds. I.E. acknowledges funding from the Spanish Ministry of Science and Innovation (Grant No. PID2022-142861NA-I00) and the Department of Education, Universities and Research of the Eusko Jaurlaritza and the University of the Basque Country UPV/EHU (Grant No. IT1527-22).